\documentclass [12pt] {article}
\usepackage{epsfig}
\begin{document}
\title{Subthreshold and near-threshold kaon and antikaon production in proton-nucleus reactions}
\author{A. V. Akindinov$^{a}$, M. M. Chumakov$^{a}$, M. M. Firoozabadi$^{b}$,\\ Yu. T. Kiselev$^{a}$, A. N. Martemyanov$^{a}$,
E. Ya. Paryev$^{c}$,\\ V. A. Sheinkman$^{a}$, Yu. V. Terekhov$^{a}$, and V. I. Ushakov$^{a}$\\
{\it  $^{a}$Institute for Theoretical and Experimental Physics,}\\
{\it Moscow 117218, Russia}\\
{\it  $^{b}$Department of Physics, Birjand University,}\\
{\it Birjand 9717914141, Iran}\\
{\it  $^{c}$Institute for Nuclear Research, Russian Academy of Sciences,}\\
{\it Moscow 117312, Russia}}


\maketitle

\begin{abstract}

The differential production cross sections of $K^+$ and $K^-$ mesons have been measured at the ITEP proton synchrotron
in $p+Be$, $p+Cu$ collisions under lab angle of 10.5$^{\circ}$, respectively, at 1.7 and 2.25, 2.4 GeV beam energies. A detailed
comparison of these data with the results of calculations within an appropriate folding model for incoherent primary
proton--nucleon, secondary pion--nucleon kaon and antikaon production processes and processes associated with the creation
of antikaons via the decay of intermediate phi mesons is given. We show that the strangeness exchange process $YN \to NNK^-$ gives
a small contribution to the antikaon yield in the kinematics of the performed experiment. We argue that in the case when
antikaon production processes are dominated by the channels with $K$$K^-$ in the final state, the cross sections of
the corresponding reactions are weakly influenced by the in-medium kaon and antikaon mean fields.
\end{abstract}

\newpage

\section*{1. Introduction}

\hspace{1.5cm} The study of the kaon and antikaon production in proton-nucleus collisions at incident energies near or below the free
nucleon--nucleon thresholds has received  considerable interest in recent years (see, for example, [1--27]). This interest has
been particularly motivated by the hope to extract from this study information about both the intrinsic properties of target
nuclei (such as high-momentum components of the nuclear wave function) and the in-medium kaon and antikaon properties
at the density of ordinary nuclei (mean-field nuclear potentials, their momentum dependencies).
Evidently, to draw the firm conclusions on these properties from proton-induced reactions
it is of principal importance to disentangle reliably the underlying reaction mechanisms,
since if the ones are clearly identified the above quantities can be fixed by comparison with the data.
Therefore, we will focus in this study on the different elementary processes that lead to the kaon and antikaon production at
beam energies close to their production thresholds in free NN collisions (1.58 GeV for $K^+$ creation and 2.5 GeV for $K^-$
creation in these collisions).

   Until now, a lot of data on total [1] and double differential [2--8, 18] $K^+$ production cross sections in proton--nucleus collisions
at bombarding energies between 0.8 and 2.9 GeV has been collected. Whereas the experimental $K^+$ spectra have been taken
[2, 3, 4, 18] mostly at laboratory emission angles $\ge 15^{\circ}$, only in two experiments [5, 6--8] the kaon yields have been obtained at
forward angles ($<15^{\circ}$). In the experiment [6--8], $K^+$ mesons were measured at laboratory angles $<12^{\circ}$ and for
momenta in the restricted range of 150--600 MeV/c, while in the experiment [5], kaons with an emission angle of 10.5$^{\circ}$ and a
fixed momentum of 1.28 GeV/c were detected. In this respect, it is highly desirable and useful to measure differential $K^+$
production cross sections in $pA$ interactions at beam energies below or close to the nucleon--nucleon threshold at forward
emission angles and for the momenta ranging from $\sim$ 0.6 to $\sim$ 1.0--1.3 GeV/c,
since such data in combination with the available
differential data may help to shed light on the mechanism of subthreshold and near-threshold kaon creation
in these interactions.

       As far as the production of $K^-$ mesons on nuclei by protons at incident energies near or below NN-threshold is concerned,
only a few data exist nowadays [14--18]. $K^-$ production has been investigated at KEK for C and Cu targets at initial energies
between 3.5 and 5.0 GeV by inclusively measuring $K^-$ at 5.1$^{\circ}$ for momentum of 1.5 GeV/c [14]. Energy dependence
of the inclusive invariant cross section for the production of antikaons with momentum of  0.8 GeV/c at an angle of 24$^{\circ}$
by protons on carbon nuclei has been measured at JINR in the region of beam energies of 3.6--8 GeV [15]. An analogous dependence,
but for $K^-$ mesons with momentum of 1.28 GeV/c and an emission angle of 10.5$^{\circ}$, was obtained at ITEP using proton
energies from 2.25 to 2.92 GeV and targets Be, Al, Cu [16, 17]. The $K^-$ momentum spectra from reactions $p+A \to {K^-}+X$
with A$=$C and Au at laboratory angles from 40$^{\circ}$ to 56$^{\circ}$ and bombarding energies of 2.5 and 3.5 GeV have been
measured at GSI by the KaoS Collaboration [18]. So, up to now, there exist no data on the antikaon momentum spectra from
proton--nucleus collisions at beam energies below the absolute threshold for free NN interactions. It should be pointed out
that the near-threshold $K^-$ production in $pA$ reactions is studied presently at the accelerator COSY [28].

     This paper presents the $K^+$ and $K^-$ momentum distributions at an angle of 10.5$^{\circ}$ in the lab frame on Be, Cu target
nuclei measured at ITEP at initial proton kinetic energies of 1.7 and 2.25, 2.4 GeV, respectively, as well as their analysis within an
appropriate folding model for incoherent primary and secondary elementary (anti)kaon production processes.
     The inclusive $K^+$ meson production
will be analyzed by us with respect to the commonly used [3, 5, 9, 10, 12, 13, 24--27]  one-step
($pN \to K^+YN$, $Y=\Lambda, \Sigma$) and two-step ($pN \to {\pi}X$, ${\pi}N \to K^+Y$) incoherent production processes.
 The inclusive rarer $K^-$ meson creation will be described not only by the considering the conventional direct
($pN \to NNKK^-$) and two-step ($pN \to {\pi}X$, ${\pi}N \to NKK^-$) antikaon production processes [17, 19--23],
but also with accounting for the new elementary reaction channels: $pN \to pN{\phi}$, $pn \to d{\phi}$, $\phi \to K^+K^-$; non-${\phi}$
$pn \to dK^+K^-$. It should be noted that the
total cross sections of the elementary reactions $pp \to pp{\phi}$, $pn \to d{\phi}$, non-${\phi}$
$pn \to dK^+K^-$, needed for our analysis, have been recently measured in the threshold region by the ANKE-at-COSY
Collaboration [29, 30, 31].
In addition to the considered by us processes, $K^-$ mesons can be produced also in the two-step reactions $pN \to K^+YN$ followed by $YN \to NNK^-$ or $Y\pi \to NK^-$. In Ref. [18,21] was found that in proton-nucleus collisions at 2.5 GeV beam energy the contribution of the above strangeness exchange processes to the cross section for antikaon production  on heavy targets at large angles is essential. We show that the strength of this channel depends on the kinematical conditions of the particular experiment and the strangeness exchange processes appears to be unimportant in the kinematics of the present experiment.

The processes of the kaon and antikaon production from nuclei can also be influenced by their nuclear self-energies [10, 20, 32].
 Since the $K^+$ attractive nuclear potential manifests itself only in the low momentum range less than 0.3-0.4 GeV/c [7,8,9] we shall neglect it in the calculation of the production of kaons with momenta of more than 0.6 GeV/c in the $pN \to K^+YN (Y=\Lambda,\Sigma)$ reactions in the near threshold and above threshold proton energy range of 1.5 - 2.0 GeV. The role of the $K^+$ potential becomes important at deep subthreshold energies [10]. We argue below that the cross sections of the reactions with $KK^-$ in the final state like $pN \to NNKK^-$, $pn \to dKK^-$, ${\pi}N \to NKK^-$; $pN \to pN{\phi}$, $pn \to d{\phi}$, $\phi \to K^+K^-$ have only low sensitivity to the in-medium kaon and antikaon mean fields because they act in the opposite directions.

\section*{2. Experiment}

\hspace{1.5cm} The experiment was carried out  with internal proton beam of the 10 GeV ITEP synchrotron irradiating Be and
Cu strip targets of 50--100 micron thick. Initial proton energy was known within the accuracy of 5 MeV and was controlled by
permanent measurement of the accelerating frequency. Secondary particles produced in the momentum range from 0.6 to 1.3 GeV/c
were detected at  fixed emission angle of 10.5$^{\circ}$ in lab frame by Focusing Hadron Spectrometer (FHS) consisting of two
bending dipole and two pairs of quadrupole magnets. Momentum and angular acceptance of the magnetic channel are
${\Delta}p/p=\pm$ 1\% and 0.8 msr, correspondingly. Two multiwire proportional chambers located at the second focus of
magnetic system served for monitoring of the beam position during the data taking. The particle identification system included
the differential Cherenkov counter [33] and two-stage TOF system based on the scintillation counters.

     The kaon and antikaon selection in the momentum range from 0.6 to 1 GeV/c was performed by using the quartz radiator of 4$\times$6
cm$^{2}$ size and 1.6 cm thick with the refraction index $n=1.47$ at $\lambda=400$ nm. The photons emitted by the kaons in
radiator were detected by the remotely controlled ring consisting of 12 photomultipliers FEU-130. For each detected momentum
the differential  Cherenkov counter was adjusted, using the mechanisms for lengthwise and angular PM  movement, in order to
achieve the highest possible efficiencies. The kaon selection criterion incorporated the requirement that 6 PMs of the ring were
fired. The data on a larger number of fired photomultipliers (8, 10 and 12) were also available. The identification
reliability increased with an increasing number of operated PMs, although, in this case, the detection efficiency decreased.
Depending on the background level, various data sets with different number of fired photomultipliers were used in further analysis.
Kaon detection efficiency varied from 65\% to 95\% depending on the number of operated PMs; this efficiency was periodically
tested in special measurements using the protons with the same velocity.
The  Cherenkov counter velocity resolution was equal to 0.02. The totally reflected photons produced by pions reached the top and
bottom surfaces of the radiator and then were detected by two optically coupled photomultipliers FEU-115. The signals from these
photomultipliers were used to suppress the pion background with the anticoincidence channel. The pion rejection ability of 100 was
reached in the 0.5--1.0 GeV/c kaon momentum range. At higher momenta the $K^+$ and $K^-$ selection was performed with Cherenkov counter used in our previous measurements [5,17].
    The time-of-flight measurements were performed at two bases of 11 and 17 meters length. Two photomultipliers XP2020 mounted on
both sides of each scintillator were used for the TOF measurements and provided the mean-timer signals. The time resolution of the
system did not exceed 300 ps (FWHM). Our trigger for the hardware antikaon selection includes the signals from six photomultipliers of
Cherenkov counters, the signals from TOF system and absence of the anticoincidence signals. The antikaon identification was quite
reliable up to the value of $K^-/{\pi}^-$ ratio equal to 1/2$\cdot$10$^6$ at the downstream of the TOF counter. The $K^+$ and $K^-$  misidentification
was less than 5\% for the most of data and did not exceed 10\% at the lowest incident proton energy of 2.25 GeV.

Secondary (anti)kaons and pions were detected simultaneously. The measured meson yields were corrected for the losses due to their nuclear interactions and multiple scattering in the material of the spectrometer, detector efficiencies and meson in-flight decay at the distance of 30 meters from the production target to the second focus of the magnetic channel. The measurements of the $K/\pi$ ratios covered the momentum range from 0.6 to 1.3 GeV/c. Absolute values of the $K^+$ and $K^-$ production cross sections on different nuclear targets were normalized to the corresponding ${\pi}^+$ and ${\pi}^-$ cross sections which were determined in special measurements in which the pion flux was measured simultaneously with the flux of the protons traversing the sandwich type targets made of thin Al foil and Be or Cu foils. The determination of the proton flux was performed by measurement of induced $\gamma$-activity in the reaction $p + ^{27}Al \to ^{24}Na^* + X$. The cross section of this reaction is known with accuracy of 6{\%}. The resulting uncertainty of the absolute normalization of the invariant (anti)kaons cross sections was estimated as 20\%.

        The $K^+$ production data were taken at initial proton kinetic energy of 1.7 GeV, while the $K^-$ creation data
were obtained at two beam energies of 2.25 and 2.4 GeV.
All cross sections presented in Tables 1 and 2 do not include the errors of the absolute normalization of the data.
{\small
\begin{center}
\centerline{ }
\centerline{ Table 1. Invariant cross sections [GeV\mbox{{$\mu$}b}/(GeV/c)$^3$] for $K^+$ production}
\centerline{ at kinetic proton energy of 1.7 GeV and at a lab angle of 10.5$^{\circ}$}
\centerline{ }
\begin{tabular}[b]{|c|c|c|c|c|c|c|c|}\hline
         &                       &           &          &          &         &         &      \\
 p, GeV/c&      0.675 & 0.806 & 0.940 & 1.05 & 1.15 & 1.28 & \\
\hline
         &                       &          &         &      &     &    &   \\
$Ed^3{\sigma}/d^3p$& 121$\pm$18& 104$\pm$16& 43.6$\pm$6.5& 10.7$\pm$1.6  & 2.80$\pm$0.42 &0.072$\pm$0.018 & Be \\
\hline
         &                       &          &         &     &     &      &    \\
$Ed^3{\sigma}/d^3p$& 308$\pm$46& 212$\pm$32& 87.7$\pm$13.2& 33.3$\pm$5.0& 8.41$\pm$1.26  & 0.34$\pm$0.09  & Cu \\
\hline
\end{tabular}
\end{center}
}
{\small
\begin{center}
\centerline{ }
\centerline{ Table 2. Invariant cross sections [GeV\mbox{nb}/(GeV/c)$^3$] for $K^-$ production}
\centerline{ at kinetic proton energy $\epsilon_0$=2.25, 2.4 GeV and at a lab angle of 10.5$^{\circ}$}
\centerline{ }
\begin{tabular}[b]{|c|c|c|c|c|c|c|c|}\hline
         &                       &           &          &          &         &         &      \\
 p, GeV/c&      0.640 & 0.807 & 0.940 & 1.05 & 1.14 & 1.28 &     \\
\hline
         &                       &          &            &         &         &         &   \\
$Ed^3{\sigma}/d^3p$\\$\epsilon_0=2.25$ GeV
& 3200$\pm$560 & 1600$\pm$240 & 480$\pm$80 & 120$\pm$18  & 40$\pm$6 & 4.7$\pm$1.2 & Be \\
\hline
         &                       &          &            &          &         &         &   \\
$Ed^3{\sigma}/d^3p$\\$\epsilon_0=2.4$ GeV & 8530$\pm$1500 & 4555$\pm$700 & 2216$\pm$380 & 660$\pm$100 & 252$\pm$40 & 42$\pm$6 & Be \\
\hline
         &                       &          &            &           &         &         &     \\
$Ed^3{\sigma}/d^3p$\\$\epsilon_0=2.4$ GeV & 39400$\pm$8500 & 18300$\pm$2800 & 6475$\pm$900 & 2800$\pm$440 & 921$\pm$150 & 197$\pm$31 & Cu \\
\hline
\end{tabular}
\end{center}
}

In Fig. 1 the measured $K^+$ and $K^-$ cross sections on the beryllium target are presented as a function of the radial Feynman variable $X^{R}_{F}$.  The variable $X^{R}_{F} =P^*/P^{*}_{max} =\sqrt{(P_{l}^{*2}+P_{t}^{*2})}/{P^{*}_{max}}$,
where $P_{l}^{*}$ and $P_{t}^{*}$ are longitudinal
 and transverse components of the meson momentum in the proton-nucleus center-of-mass system and $P^{*}_{max}$ stands for its maximum momentum. It is seen that the cross sections for the kaon and antikaon production at near threshold and especially subthreshold collision energies drastically decrease with
 $X^{R}_{F}$. High momentum parts of the meson spectra for $X^{R}_{F} >$ 0.5 follow the dependence $Ed^{3}\sigma/d^{3}p \sim (1-X^{R}_{F})^\delta$ with the value of exponent ${\delta}>$ 4.

\section*{3. Analysis of the data. The model and inputs}

\section*{3.1. Direct $K^+$ and $K^-$ production mechanisms}

\hspace{1.5cm} The direct non-resonant production of $K^+$ and $K^-$ mesons in $pA$ collisions at incident energies of our
interest (up to 2.4 GeV) can occur due to the Fermi motion of nuclear nucleons in the following elementary processes with the
lowest thresholds:
\begin{equation}
p+N \to K^++\Lambda+N,
\end{equation}
\begin{equation}
p+N \to K^++\Sigma+N;
\end{equation}
\begin{equation}
p+N \to N+N+K+K^-,
\end{equation}
\begin{equation}
p+n \to d+{K^+}+K^-,
\end{equation}
where $ K$ stands for $K^+$ or $K^0$ for the specific isospin channel. The invariant inclusive cross sections for the production on nucleus with atomic mass number
$A$ $K^+$ and $K^-$ mesons with the total energies $E_{K^+}$ and $E_{K^-}$ and momenta  ${\bf p}_{K^+}$ and
 ${\bf p}_{K^-}$ at small laboratory emission angles from the primary proton--induced reaction channels (1)--(4) can be presented as follows [17]
\footnote{We can neglect in the energy domain of our interest the additional contribution of the $K^+$ production processes (3), (4)
to the kaon creation cross section (5) due to their larger production thresholds in $pN$ collisions.
}
:
\begin{equation}
E_{K^+}\frac{d\sigma_{pA\to K^+X}^{({\rm prim})}({\bf p}_0)}
{d{\bf p}_{K^+}}=
I_{K^+}[A]\times
\end{equation}
$$
\left\{\left<E_{K^+}\frac{d\sigma_{pN\to K^+\Lambda N}({\bf p}_0,{\bf p}_{K^+})}
{d{\bf p}_{K^+}}\right>+
\left<E_{K^+}\frac{d\sigma_{pN\to K^+\Sigma N}({\bf p}_0,{\bf p}_{K^+})}
{d{\bf p}_{K^+}}\right>\right\},
$$
\begin{equation}
E_{K^-}\frac{d\sigma_{pA\to K^-X}^{({\rm prim})}({\bf p}_0)}
{d{\bf p}_{K^-}}=
I_{K^-}[A]\times
\end{equation}
$$
\left\{\left<E_{K^-}\frac{d\sigma_{pN\to NNKK^-}({\bf p}_0,{\bf p}_{K^-})}
{d{\bf p}_{K^-}}\right>+\frac{N}{A}
\left<E_{K^-}\frac{d\sigma_{pn\to dK^+K^-}({\bf p}_0,{\bf p}_{K^-})}
{d{\bf p}_{K^-}}\right>\right\},
$$
where
\begin{equation}
I_{h}[A]=2{\pi}A\int\limits_{0}^{\infty}r_{\bot}dr_{\bot}
\int\limits_{-\infty}^{+\infty}dz
\rho(\sqrt{r_{\bot}^2+z^2})\times
\end{equation}
$$
\times
\exp{\left[-\sigma_{pN}^{{\rm in}}A\int\limits_{-\infty}^{z}
\rho(\sqrt{r_{\bot}^2+x^2})dx-\sigma_{hN}^{{\rm tot}}(p_h)A
\int\limits_{z}^{+\infty}\rho(\sqrt{r_{\bot}^2+x^2})dx\right]},
$$
\begin{equation}
\sigma_{hN}^{{\rm tot}}(p_h)=(Z/A)\sigma_{hp}^{{\rm tot}}(p_h)+(N/A)\sigma_{hn}^{{\rm tot}}(p_h)
\end{equation}
and
\begin{equation}
\left<E_{K^+}\frac{d\sigma_{pN\to K^+YN}({\bf p}_0,{\bf p}_{K^+})}
{d{\bf p}_{K^+}}\right>=
\int
n({\bf p}_t)d{\bf p}_t\times
\end{equation}
$$
\left[E_{K^+}
\frac
{d\sigma_{pN\to K^+YN}(\sqrt{s}, {\bf p}_{K^+})}
{d{\bf p}_{K^+}}\right],
$$
\begin{equation}
\left<E_{K^-}\frac{d\sigma_{pN\to NNKK^-(pn\to dK^+K^-)}({\bf p}_0,{\bf p}_{K^-})}
{d{\bf p}_{K^-}}\right>=
\int
n({\bf p}_t)d{\bf p}_t\times
\end{equation}
$$
\left[E_{K^-}
\frac
{d\sigma_{pN\to NNKK^-(pn\to dK^+K^-)}(\sqrt{s}, {\bf p}_{K^-})}
{d{\bf p}_{K^-}}\right].
$$
Here,
$E_{K^+}d\sigma_{pN\to K^+YN}(\sqrt{s},{\bf p}_{K^+}) /d{\bf p}_{K^+}$ and
$E_{K^-}d\sigma_{pN\to NNK{K^-}}(\sqrt{s},{\bf p}_{K^-}) /d{\bf p}_{K^-}$,\\
$E_{K^-}d\sigma_{pn\to dK^+{K^-}}(\sqrt{s},{\bf p}_{K^-}) /d{\bf p}_{K^-}$
are the off-shell invariant inclusive
cross sections for $K^+$ and $K^-$ production in reactions (1), (2) and (3), (4), respectively;
$\rho({\bf r})$ and
$n({\bf p}_t)$ are the density and internal nucleon momentum distribution normalized to unity;
${\bf p}_t$ is the momentum of the struck target nucleon just before the collision;
$\sigma_{pN}^{{\rm in}}$ and $\sigma_{hp(hn)}^{{\rm tot}}$
are the inelastic and total cross sections
of free $pN$ and $hp$ ($hn$) interactions;
$Z$ and $N$ are the numbers of protons and neutrons in the target nucleus;
${\bf p}_0$ is the momentum of the initial proton; $s$ is the $pN$ center-of-mass energy squared.
The expression for $s$ is:
\begin{equation}
  s=(E_{0}+E_t)^2-({\bf p}_{0}+{\bf p}_t)^2,
\end{equation}
where $E_0$ and $E_t$ are the total energies of the projectile proton and the struck nucleon, respectively.
The total energy $E_t$ of the off-shell intranuclear nucleon is related to its momentum ${\bf p}_t$ within
the considered model as follows [5, 17]:
\begin{equation}
  E_t=m_{N}-\frac{{\bf p}_t^2}{2M_{A-1}}-\epsilon.
\end{equation}
Here, $m_N$ is the nucleon mass, $M_{A-1}$ is the mass of the recoiling target nucleus in its ground state, and
the separation energy $\epsilon$ is taken as equal to 2 MeV.

    In eqs. (5) and (6) it is assumed that the corresponding $K^+$ and $K^-$ meson production cross sections in $pp$
and $pn$ interactions are the same [3, 12, 17, 19, 22]. The quantities $I_{K^+}[A]$ and $I_{K^-}[A]$ in these equations
describe the attenuation of the proton beam as well as the distortion of the outgoing kaons and antikaons in their ways out
of the nucleus. When calculating them according to the eq. (7), we use $\sigma_{pN}^{{\rm in}}=30$ mb for the considered
projectile proton energies [12, 22], we adopt also $\sigma_{K^+p}^{{\rm tot}}=\sigma_{K^+n}^{{\rm tot}}=12$ mb for all kaon
momenta of our interest [34], and for the antikaon--nucleon total cross sections $\sigma_{K^-p}^{{\rm tot}}(p_{K^-})$,
$\sigma_{K^-n}^{{\rm tot}}(p_{K^-})$ as functions of the $K^-$ momentum $p_{K^-}$ we employ the respective parametrizations
suggested in [35]
\footnote{At antikaon momenta $p_{K^-} \le 0.63$ GeV/c we use the following relation
$\sigma_{K^-n}^{{\rm tot}}(p_{K^-})=0.8\sigma_{K^-p}^{{\rm tot}}(p_{K^-})$ [36].}
.
       The nuclear density $\rho({\bf r})$ for the Be and C target nuclei of interest is given by
the harmonic oscillator distribution, whereas the density for the Cu nucleus is specified by the Woods-Saxon distribution [22, 37].

   In the expressions (9) and (10), entering into the eqs. (5) and (6), the differential cross sections for the production of $K^+$ and
$K^-$ mesons in a collision of a proton with an intranuclear nucleon , having momentum ${\bf p}_t$, are folded with the internal
nucleon momentum distribution $n({\bf p}_t)$. This distribution was previously determined by fitting the $K^+$ production cross
sections in $pA$ reactions at near-threshold and subthreshold initial energies [5]. It was assumed to be in the form
\begin{equation}
  n({\bf p}_t)=\frac{1}{(2\pi)^{3/2}(1+h)}
\left[\frac{1}{\sigma_{1}^{3}}
\exp{(-p_{t}^{2}/2\sigma_{1}^{2})}+
\frac{h}{\sigma_{2}^{3}}\exp{(-p_{t}^{2}/2\sigma_{2}^{2})}\right].
\end{equation}
 Here, $h=0.12$, $\sigma_{2}=220$ MeV/c for the Be, C and Cu target nuclei of present interest; $\sigma_{1}=132$ and 146 MeV/c
for the Be, C and Cu nuclei, respectively.

        Let us now specify the off-shell invariant inclusive cross sections
$E_{K^+}d\sigma_{pN\to K^+YN}(\sqrt{s},{\bf p}_{K^+}) /d{\bf p}_{K^+}$\\ and
$E_{K^-}d\sigma_{pN\to NNKK^-}(\sqrt{s},{\bf p}_{K^-}) /d{\bf p}_{K^-}$,
$E_{K^-}d\sigma_{pn\to dK^+K^-}(\sqrt{s},{\bf p}_{K^-}) /d{\bf p}_{K^-}$
for $K^+$ and $K^-$ production in the reactions (1), (2) and (3), (4), entering into eqs. (5), (9) and (6), (10), respectively.
Following refs. [17, 22, 23], we assume that these
cross sections are equivalent to the respective on-shell cross sections calculated for the off-shell kinematics of the elementary
processes  (1)--(4) at the same collision energy $\sqrt{s}$.

In our approach the invariant inclusive cross sections for $K^+$ production in the reactions (1), (2) have been
described by the three-body phase-space calculations corrected for the FSI effects between the hyperons and nucleons\footnote{The validity of the incoporation of the elementary FSI effects into the analisys of $pA$ reactions is an open question. Since in the subthreshold  and near-threshold energy region the outgoing particles are mainly emitted in the forward directions close to each other with small relative momenta and their momenta relative to the target system essentially greater then the Fermi momentum, one may hope that the bare FSI is not drastically suppressed in nuclei. Below we present the results of the calculations with and without including FSI effects.}
following the corresponding Watson-Migdal theory [38, 39] and adopting the inverse Jost-function approximation [23] for the
FSI amplitude $M_{\rm FSI}^{YN}$:
\begin{equation}
E_{K^{+}}\frac{d\sigma_{pN\rightarrow K^+YN}(\sqrt{s},{\bf p}_{K^{+}})}
{d{\bf p}_{K^{+}}}
={\frac{{\pi}}{4}}
{\frac{\sigma_{pN\rightarrow K^+YN}({\sqrt{s}})}
{I_{3}(s,m_K,m_{Y},m_{N},F_{\rm FSI}^{YN})}}\times
\end{equation}
$$\times
{\frac{\lambda(s_{YN},m_{Y}^{2},m_{N}^{2})}{s_{YN}}}F_{\rm FSI}^{YN}(s_{YN}),
$$
where
\begin{equation}
I_{3}(s,m_K,m_{Y},m_{N},F_{\rm FSI}^{YN})=(\frac{{\pi}}{2})^2
\int\limits_{(m_{Y}+m_{N})^2}^{({\sqrt{s}}-m_K)^2}
{\frac{\lambda(s_{YN},m_{Y}^{2},m_{N}^{2})}{s_{YN}}}\times
\end{equation}
$$\times
{\frac{\lambda(s,s_{YN},m_{K}^{2})}{s}F_{\rm FSI}^{YN}(s_{YN})\,ds_{YN}},
$$
\begin{equation}
\lambda(x,y,z)=\sqrt{{\left[x-({\sqrt{y}}+{\sqrt{z}})^2\right]}{\left[x-
({\sqrt{y}}-{\sqrt{z}})^2\right]}},
\end{equation}
\begin{equation}
s_{YN}=s+m_{K}^{2}-2(E_{0}+E_t)E_{K^{+}}+
2({\bf p}_{0}+{\bf p}_t){\bf p}_{K^{+}},
\end{equation}
\begin{equation}
             E_{0}=\sqrt{{\bf p}_{0}^{2}+m_{N}^{2}}, \,\,\,\,E_{K^+}=\sqrt{{\bf p}_{K^+}^{2}+m_{K}^{2}}.
\end{equation}
 Here, $\sigma_{pN\rightarrow K^+YN}$ are the total cross sections for $K^+$ production in reactions (1), (2) taken from [13];
 $m_{K}$ and $m_{Y}$ are the masses in free space of a kaon and a $Y$ hyperon ($\Lambda$ or $\Sigma$), respectively;
$F_{\rm FSI}^{YN}$ is the so-called $YN$-FSI enhancement factor. This factor is given by [23]:
\begin{equation}
F_{{\rm FSI}}^{YN}(s_{YN})=\left|M_{\rm FSI}^{YN}(q_{YN})\right|^2=
\frac{q_{YN}^2+{\alpha}_{YN}^2}{q_{YN}^2+{\beta}_{YN}^2}
\end{equation}
and
\begin{equation}
q_{YN}=\frac{1}{2\sqrt{s_{YN}}}\lambda(s_{YN},m_{Y}^{2},m_{N}^{2}).
\end{equation}
The values of parameters $\alpha_{{\Lambda}p}$=255.55 MeV/c and $\beta_{{\Lambda}p}$=-83.38 MeV/c were calculated taking into account average $S$-wave ${\Lambda}p$ scattering length $\bar{a}$ and effective range $\bar{r}$ from [40].

     Further, taking into consideration the main
\footnote{As is well known, the $KN$ interaction is rather weak compared to the strong $NN$ interaction.}
FSI effects among the final nucleons
\footnote{Protons in the exit channel in line with the assumption that the $K^-$ meson production cross sections in $pp$ and $pn$
interactions are the same.}
participating in the primary proton--induced reaction channel $pN \to NNKK^-$ on the same footing as that adopted in calculating
the $K^+$ creation cross sections (14) from the elementary processes (1), (2) as well as following the four-body phase-space model,
we can represent the invariant inclusive cross section for $K^-$ production in this channel as follows (see, also, [23]):
\begin{equation}
E_{K^-}
\frac{d\sigma_{pN\to NNKK^-}(\sqrt{s},{\bf p}_{K^-})}
{d{\bf p}_{K^-}}=
\sigma_{pN \to NNKK^-}({\sqrt{s}})
f_{4}^{{\rm FSI}}(s,{\bf p}_{K^-}),
\end{equation}
where
\begin{equation}
f_{4}^{{\rm FSI}}(s,{\bf p}_{K^-}) =
I_{3}(s_{NNK},m_{K},m_{N},m_{N},F_{{\rm FSI}}^{NN})/
\left[2I_{4}(s,m_{K},m_{K},m_{N},m_{N},
F_{{\rm FSI}}^{NN})\right],
\end{equation}
\begin{equation}
 I_{4}(s,m_{K},m_{K},m_{N},m_{N},F_{{\rm FSI}}^{NN})=
(\frac{{\pi}}{2})\int \limits_{4m_{N}^{2}}^{(\sqrt{s}-2m_{K})^2}
{\frac{\lambda(s_{NN},m_{N}^{2},m_{N}^{2})}{s_{NN}}}
F_{{\rm FSI}}^{NN}(s_{NN})\times
\end{equation}
$$\times
 I_{3}(s,m_{K},\sqrt{s_{NN}},m_{K},F_{{\rm FSI}}^{NN}=1)ds_{NN},
$$
\begin{equation}
s_{NNK}=s+m_{K}^{2}-2(E_{0}+E_t)E_{K^{-}}+
2({\bf p}_{0}+{\bf p}_t){\bf p}_{K^{-}}
\end{equation}
and
\begin{equation}
             E_{K^-}=\sqrt{{\bf p}_{K^-}^{2}+m_{K}^{2}}.
\end{equation}
 Here, $\sigma_{pN\rightarrow NNKK^-}$ is the total cross section for the nonresonant (non-$\phi$) $K^-$ production
in reaction (3). The $NN$-FSI-modified three-body phase-space volume
$I_{3}(s_{NNK},m_{K},m_{N},m_{N},F_{{\rm FSI}}^{NN})$ and the $NN$-FSI enhancement factor $F_{\rm FSI}^{NN}$,
appearing in (22), (23), are described by the formulas (15), (19), (20), respectively, in which one has to make the following
substitutions: $s \to s_{NNK}$, $Y \to N$. The parameters ${\alpha}_{NN}$ and  ${\beta}_{NN}$, which govern the factor $F^{NN}_{FSI}$, are specified as  ${\alpha}_{NN}$=164.3 MeV/c  and ${\beta}_{NN}$=-21.9 MeV/c.

For the free total cross section
$\sigma_{pN \to NNKK^-}$ we have used the following parametrization:
\begin{equation}
\sigma_{pN \to NNKK^-}({\sqrt{s}})=\left\{
\begin{array}{ll}
	70\left(1-\frac{s_{th}}{s}\right)^{2.5}~[{\rm {\mu}b}]
	&\mbox{for $0<\sqrt{s}-\sqrt{s_{th}}\le 0.114~{\rm GeV}$}, \\
	&\\
                   270~\left(1-\frac{s_{th}}{s}\right)^{3}\left(\frac{s_{th}}{s}\right)^{0.8}~[{\rm {\mu}b}]
	&\mbox{for $\sqrt{s}-\sqrt{s_{th}}> 0.114~{\rm GeV}$},
\end{array}
\right.	
\end{equation}
where $\sqrt{s_{th}}=2(m_N+m_{K})$ is the threshold energy for the $pN \to NNKK^-$ reaction.
Its first low-excess energy part accounts for the world near-threshold experimental data for the total cross section
for the production of the non-$\phi$ component in the $pp \to ppK^+K^-$ reaction reported in [29]. The second
high-excess energy part of (26) represents the respective parametrization of the calculated exclusive total cross section
of antikaon production in nucleon--nucleon interaction within the pion and kaon exchange model [41], that has been reduced by
the factor of about 1.1 to match the first one at 114 MeV excess energy.

      Finally, neglecting the weak final-state interaction between $K^+$ meson and deuteron, produced in the reaction (4) [31, 42],
as well as using the pure three-body phase-space model, we can readily get the following expression for the invariant inclusive
cross section for nonresonant $K^-$ creation in this reaction (cf. eqs. (14), (15)):
\begin{equation}
E_{K^{-}}\frac{d\sigma_{pn\rightarrow dK^+K^-}(\sqrt{s},{\bf p}_{K^{-}})}
{d{\bf p}_{K^{-}}}
={\frac{{\pi}}{4}}
{\frac{\sigma_{pn\rightarrow dK^+K^-}({\sqrt{s}})}
{I_{3}(s,m_K,m_{d},m_{K})}}\times
\end{equation}
$$\times
{\frac{\lambda(s_{Kd},m_{K}^{2},m_{d}^{2})}{s_{Kd}}},
$$
where
\begin{equation}
I_{3}(s,m_K,m_{d},m_{K})=(\frac{{\pi}}{2})^2
\int\limits_{(m_{d}+m_{K})^2}^{({\sqrt{s}}-m_K)^2}
{\frac{\lambda(s_{Kd},m_{d}^{2},m_{K}^{2})}{s_{Kd}}}\times
\end{equation}
$$\times
{\frac{\lambda(s,s_{Kd},m_{K}^{2})}{s}\,ds_{Kd}},
$$
\begin{equation}
s_{Kd}=s+m_{K}^{2}-2(E_{0}+E_t)E_{K^{-}}+
2({\bf p}_{0}+{\bf p}_t){\bf p}_{K^{-}}.
\end{equation}
Here, $\sigma_{pn\rightarrow dK^+K^-}$ is the total cross section for the non-$\phi$ component of the
$pn \to dK^+K^-$ reaction; $m_d$ is the mass of a deuteron. The authors of [31] have fitted their own experimental data
on the total cross section $\sigma_{pn \rightarrow dK^+K^-}$ close to threshold by the following expression:
\begin{equation}
\sigma_{pn\rightarrow dK^+K^-}(\sqrt{s})=\frac{1}{4}(\sigma_{0}+\sigma_1),
\end{equation}
\begin{equation}
\sigma_{0}=A_0{\epsilon}^2/D,\,\,\sigma_{1}=A_1{\epsilon}^3(D+\frac{1}{2}\epsilon/{\epsilon}^*)/D^2,\,\,
D=(1+\sqrt{1+\epsilon/{\epsilon}^*})^2,
\end{equation}
where $\epsilon=\sqrt{s}-(m_d+2m_K)$, ${\epsilon}^*=20$ MeV, $A_0=127~{\rm {\mu}b}$/GeV$^2$, and
$A_1=1800~{\rm {\mu}b}$/GeV$^3$.

\section*{3.2. Two-step $K^+$ and $K^-$ production mechanisms}

\hspace{1.5cm} At the bombarding energies
of our interest ($\le 2.4~$GeV) the following two-step $K^+$ and $K^-$ production
processes with a pion and a $\phi$-meson in an intermediate states
 can contribute to the kaon and antikaon production in $pA$ interactions [12, 13, 22, 30, 37]
\footnote{It should be pointed out that in the intermediate pion energy region of interest the $K^-$ mesons can be
produced in ${\pi}N$ collisions also by the decay of the $\phi$ meson as an intermediate state in the process
${\pi}N \to {\phi}N$, $\phi \to K^+K^-$. However, in view of the fact that the resonant ($\phi$ meson) to total $K^-$
production cross section ratio for example in ${\pi}^-p$ reactions at pion energies of interest, as showed our estimates using
the results given in [43, 44], is rather small (about 0.1), it is natural to assume, calculating the $K^-$ yields in $pA$ reactions
from the secondary channel (35), that the antikaons are produced directly in this channel.}
:
\begin{equation}
p+N_{1}\to \pi+X,
\end{equation}
\begin{equation}
\pi + N_2\to K^++\Lambda,
\end{equation}
\begin{equation}
\pi + N_2\to K^++\Sigma;
\end{equation}
\begin{equation}
\pi + N_2\to N+K+K^-,
\end{equation}
\begin{equation}
p+N \to p+N+\phi,
\end{equation}
\begin{equation}
p+n \to d+\phi,
\end{equation}
\begin{equation}
\phi \to K^++K^-.
\end{equation}
The $K^+$ and $K^-$ production cross sections for $pA$ interactions at small laboratory angles from the
secondary pion--induced reaction channels (33), (34) and (35) can be represented as follows [12, 17, 22]:
\begin{equation}
E_{K^+}\frac{d\sigma_{pA\to {K^+}X}^{({\rm sec})}
({\bf p}_0)}
{d{\bf p}_{K^+}}=\frac{I_{K^+}^{({\rm sec})}[A]}{I_{\pi}[A]}
\sum_{\pi=\pi^+,\pi^0,\pi^-}\sum_{Y=\Lambda,\Sigma}\int \limits_{4\pi}d{\bf \Omega}_{\pi}
\int \limits_{p_{\pi}^{{\rm abs,+}}}^{p_{\pi}^{{\rm lim}}
(\vartheta_{\pi})}p_{\pi}^{2}
dp_{\pi}
\frac{d\sigma_{pA\to {\pi}X}^{({\rm prim})}({\bf p}_0)}{d{\bf p}_{\pi}}\times
\end{equation}
$$
\times
\left[\frac{Z}{A}\left<E_{K^+}\frac{d\sigma_{{\pi}p\to{K^+}Y}({\bf p}_{\pi},
{\bf p}_{K^+})}{d{\bf p}_{K^+}}\right>+\frac{N}{A}\left<E_{K^+}\frac{d\sigma_{{\pi}n\to{K^+}Y}({\bf p}_{\pi},
{\bf p}_{K^+})}{d{\bf p}_{K^+}}\right>\right],
$$
\begin{equation}
E_{K^-}\frac{d\sigma_{pA\to {K^-}X}^{({\rm sec})}
({\bf p}_0)}
{d{\bf p}_{K^-}}=\frac{I_{K^-}^{({\rm sec})}[A]}{I_{\pi}[A]}
\sum_{\pi=\pi^+,\pi^0,\pi^-}\int \limits_{4\pi}d{\bf \Omega}_{\pi}
\int \limits_{p_{\pi}^{{\rm abs,-}}}^{p_{\pi}^{{\rm lim}}
(\vartheta_{\pi})}p_{\pi}^{2}
dp_{\pi}
\frac{d\sigma_{pA\to {\pi}X}^{({\rm prim})}({\bf p}_0)}{d{\bf p}_{\pi}}\times
\end{equation}
$$
\times
\left[\frac{Z}{A}\left<E_{K^-}\frac{d\sigma_{{\pi}p\to NK{K^-}}({\bf p}_{\pi},
{\bf p}_{K^-})}{d{\bf p}_{K^-}}\right>+\frac{N}{A}\left<E_{K^-}\frac{d\sigma_{{\pi}n\to NK{K^-}}({\bf p}_{\pi},
{\bf p}_{K^-})}{d{\bf p}_{K^-}}\right>\right],
$$
where
\begin{equation}
I_{h}^{({\rm sec})}[A]=2{\pi}A^2\int\limits_{0}^{+\infty}r_{\bot}dr_{\bot}
\int\limits_{-\infty}^{+\infty}dz
\rho(\sqrt{r_{\bot}^2+z^2})
\int\limits_{0}^{+\infty}dl
\rho(\sqrt{r_{\bot}^2+(z+l)^2})
\times
\end{equation}
$$
\times
\exp{\left[-\sigma_{pN}^{{\rm in}}A\int\limits_{-\infty}^{z}
\rho(\sqrt{r_{\bot}^2+x^2})dx
-\sigma_{{\pi}N}^{{\rm tot}}A\int\limits_{z}^{z+l}
\rho(\sqrt{r_{\bot}^2+x^2})dx\right]}
\times
$$
$$
\times
\exp{\left[-\sigma_{{h}N}^{{\rm tot}}(p_h)A\int\limits_{z+l}^{+\infty}
\rho(\sqrt{r_{\bot}^2+x^2})dx\right]},
$$
\begin{equation}
\left<E_{K^+}\frac{d\sigma_{{\pi}N\to {K^+}Y}({\bf p}_{\pi},
{\bf p}_{K^+})}
{d{\bf p}_{K^+}}\right>=
\int
n({\bf p}_t)d{\bf p}_t
\left[E_{K^+}\frac{d\sigma_{{\pi}N\to {K^+}Y}(\sqrt{s_1},{\bf p}_{K^+})}
{d{\bf p}_{K^+}}\right],
\end{equation}
\begin{equation}
\left<E_{K^-}\frac{d\sigma_{{\pi}N\to NK{K^-}}({\bf p}_{\pi},
{\bf p}_{K^-})}
{d{\bf p}_{K^-}}\right>=
\int
n({\bf p}_t)d{\bf p}_t
\left[E_{K^-}\frac{d\sigma_{{\pi}N\to NK{K^-}}(\sqrt{s_1},{\bf p}_{K^-})}
{d{\bf p}_{K^-}}\right];
\end{equation}
\begin{equation}
  s_1=(E_{\pi}+E_{t})^2-(p_{\pi}{\bf \Omega_{0}}+{\bf p}_{t})^2,
\end{equation}
\begin{equation}
 p_{\pi}^{{\rm lim}}(\vartheta_{\pi}) =
\frac{{\beta}_{A}p_{0}\cos{\vartheta_{\pi}}+
 (E_{0}+M_A)\sqrt{{\beta}_{A}^2-4m_{\pi}^{2}(s_{A}+
p_{0}^{2}\sin^{2}{\vartheta_{\pi}})}}{2(s_{A}+
p_{0}^{2}\sin^{2}{\vartheta_{\pi}})},
\end{equation}
\begin{equation}
 {\beta}_A=s_{A}+m_{\pi}^{2}-M_{A+1}^{2},\,\,s_A=(E_{0}+M_A)^2-p_{0}^{2},
\end{equation}
\begin{equation}
\cos{\vartheta_{\pi}}={\bf \Omega}_0{\bf \Omega}_{\pi},\,\,\,\,
{\bf \Omega}_{0}={\bf p}_{0}/p_{0},\,\,\,\,{\bf \Omega}_{\pi}={\bf p}_{\pi}/p_{\pi}.
\end{equation}
Here, $d\sigma_{pA\to {\pi}X}^{({\rm prim})}({\bf p}_0)/d{\bf p}_{\pi}$ are the
inclusive differential cross sections for pion production on nuclei at small laboratory angles and for high momenta from
the primary proton--induced reaction channel (32);
$E_{K^+}d\sigma_{{\pi}N\to {K^+}Y}(\sqrt{s_1},{\bf p}_{K^+})/d{\bf p}_{K^+}$ and
$E_{K^-}d\sigma_{{\pi}N\to NK{K^-}}(\sqrt{s_1},{\bf p}_{K^-})/d{\bf p}_{K^-}$ are
the free inclusive invariant cross sections for $K^+$ and $K^-$ production via the subprocesses (33), (34) and (35)
 calculated for the off-shell kinematics of
these subprocesses at the ${\pi}N$ center-of-mass energy $\sqrt{s_1}$;
$\sigma_{\pi N}^{{\rm tot}}$ is the total cross section of the free $\pi N$ interaction
\footnote{We use in the following calculations $\sigma_{\pi N}^{{\rm tot}}=35$ mb for all pion momenta [22].}
;
${\bf p}_{\pi}$ and $E_{\pi}$ are the momentum and total energy of a pion (which is assumed to be on-shell);
$m_{\pi}$ is the rest mass of a pion;
$p_{\pi}^{{\rm abs},+(-)}$ is the absolute threshold momentum for kaon (antikaon) production on the residual nucleus
by an intermediate pion;
$p_{\pi}^{{\rm lim}}(\vartheta_{\pi})$ is the kinematical limit for pion production
at the lab angle $\vartheta_{\pi}$ from proton-nucleus collisions; $M_A$ and $M_{A+1}$ are the ground state masses of the
initial target nucleus and nucleus containing $A+1$ nucleons.
The quantity $I_{\pi}[A]$, entering into (39), (40), is defined above by eq. (7) in which one has to make the substitution $h \to \pi$.

       The elementary $K^+$ and $K^-$ production reactions ${\pi}^+n \to K^+\Lambda$, ${\pi}^0p \to K^+\Lambda$,
${\pi}^+p \to K^+{\Sigma}^+$, ${\pi}^+n \to K^+{\Sigma}^0$,  ${\pi}^0p \to K^+{\Sigma}^0$,  ${\pi}^0n \to K^+{\Sigma}^-$,
 ${\pi}^-p \to K^+{\Sigma}^-$ and  ${\pi}^+n \to pK^+K^-$, ${\pi}^0p \to pK^+K^-$, ${\pi}^0n \to nK^+K^-$,
${\pi}^0n \to pK^0K^-$, ${\pi}^-p \to nK^+K^-$, ${\pi}^-p \to pK^0K^-$, ${\pi}^-n \to nK^0K^-$ have been included in our
calculations of the $K^+$ and $K^-$ production on nuclei. In them, the differential cross sections for kaon and antikaon creation
in these reactions have been computed following strictly the approach [12, 22].

     Another very important ingredients for the calculation of the $K^+$ and $K^-$ production cross sections
 in proton--nucleus reactions
from pion-induced reaction channels (33), (34) and (35)--the high-momentum parts of the
differential cross sections for pion production on
nuclei at small lab angles from the primary process (32)--for $^{9}$Be and $^{63}$Cu target nuclei were taken from [17, 22].

Now consider the $K^-$ production in $pA$ reactions via the phi production/decay sequences (36)--(38).
Taking into account the fact that the most of $\phi$ mesons, which are responsible for the subthreshold antikaon production in
$pA$ collisions for kinematics of our experiment, decay into $K^+$ and $K^-$ mesons essentially outside the target nuclei of
interest [37, 45]  as well as using the results given in [17, 37, 46], we get the following expression for the $K^-$ production
cross section for $pA$ interactions from these sequences:
\begin{equation}
E_{K^-}\frac{d\sigma_{pA\to {\phi}X,\, \phi \to K^+K^-}
({\bf p}_0)}
{d{\bf p}_{K^-}}=I_{\phi}[A]\left(\frac{M_{\phi}}{m_K}\right)^2BR_{\phi \to K^+K^-}(M_{\phi})
\int \frac{d{\bf \Omega}_{\phi}^{*}}{4\pi}\times
\end{equation}
$$
\times
\left[\frac{Z}{A}\left<E_{\phi}\frac{d\sigma_{pp\to pp{\phi}}({\bf p}_{0},
{\bf p}_{\phi})}{d{\bf p}_{\phi}}\right>+\frac{N}{A}\left<E_{\phi}\frac{d\sigma_{pn\to pn{\phi}}({\bf p}_{0},
{\bf p}_{\phi})}{d{\bf p}_{\phi}}\right>+
\frac{N}{A}\left<E_{\phi}\frac{d\sigma_{pn\to d{\phi}}({\bf p}_{0},
{\bf p}_{\phi})}{d{\bf p}_{\phi}}\right>\right],
$$
where
\begin{equation}
\left<E_{\phi}\frac{d\sigma_{pN\to pN{\phi}(pn\to d{\phi})}({\bf p}_{0},
{\bf p}_{\phi})}
{d{\bf p}_{\phi}}\right>=
\int
n({\bf p}_t)d{\bf p}_t
\left[E_{\phi}\frac{d\sigma_{pN\to pN{\phi}(pn\to d{\phi})}(\sqrt{s},M_{\phi},{\bf p}_{\phi})}
{d{\bf p}_{\phi}}\right].
\end{equation}
Here,
$E_{\phi}d\sigma_{pN\to pN{\phi}}(\sqrt{s},M_{\phi},{\bf p}_{\phi}) /d{\bf p}_{\phi}$ and
$E_{\phi}d\sigma_{pn\to d{\phi}}(\sqrt{s},M_{\phi},{\bf p}_{\phi}) /d{\bf p}_{\phi}$
are the off-shell
differential cross sections for $\phi$ production in reactions (36) and (37), respectively,
at the $pN$ center-of-mass energy $\sqrt{s}$ and at its pole mass
\footnote{Since a $\phi$-meson is a narrow resonance.}
$M_{\phi}$; ${\bf p}_{\phi}$ and $E_{\phi}$ are the momentum and total energy of a $\phi$ meson
($E_{\phi}=\sqrt{{\bf p}_{\phi}^{2}+M_{\phi}^{2}}$); ${\bf \Omega}_{\phi}^{*}$ is its solid angle
in the rest frame of the detected antikaon. The quantity $I_{\phi}[A]$ in (48) is defined above by eq. (7)
in which one has to make the substitution $h \to \phi$. For the total ${\phi}N$ cross section
$\sigma_{\phi N}^{{\rm tot}}$ appearing in this quantity we have used the value of 8.3 mb [47].

       In our calculations of the $K^-$ meson production on nuclei the differential cross sections\\
$E_{\phi}d\sigma_{pN\to pN{\phi}}(\sqrt{s},M_{\phi},{\bf p}_{\phi}) /d{\bf p}_{\phi}$ and
$E_{\phi}d\sigma_{pn\to d{\phi}}(\sqrt{s},M_{\phi},{\bf p}_{\phi}) /d{\bf p}_{\phi}$, entering into eqs. (48), (49),
have been computed within the approach [45]. In doing so, the corresponding parametrization for the free total cross
section ${\sigma}_{pp \to pp{\phi}}(\sqrt{s})$ of the $pp \to pp{\phi}$ reaction presented in [45] has been reduced
by the factor of 1.33 in line with the new set of data [29] for this reaction in the threshold region and the cross section
ratio ${\sigma}_{pn \to pn{\phi}}/{\sigma}_{pp \to pp{\phi}}$ has been employed in the excess-energy-dependent form
from [48].

\section*{3.3. Comparison with the experimental data}

\hspace{1.5cm} At first, we will concentrate on the results of our calculations for the production of $K^+$
mesons from $p$Be, $p$C and $p$Cu reactions at the various bombarding energies.

              Figures 2 and 3 show a comparison of the experimental data from the present experiment
for the invariant differential cross sections for the production of $K^+$ mesons, respectively, in $p+$Be and $p+$Cu interactions
at the laboratory angle of 10.5$^{\circ}$ and for the initial proton energy of 1.7 GeV
with the calculated ones according to (5), (39).
It is seen that the pion--induced $K^+$ creation processes (33), (34) do not
dominate in the kaon production in $p^{9}$Be collisions at all kaon momenta and the main contribution to the kaon yield here
comes from the direct $K^+$ production mechanism in both considered scenarios with and without including the FSI effects
between the hyperons and nucleons produced via this mechanism. In the case of $p^{63}$Cu interactions the direct
mechanism clearly dominates the $K^+$ production only at high kaon momenta
\footnote{This is consistent with the previous findings of [5, 13] about the role played by the direct $K^+$ production mechanism
in the "hard" kaon creation in $pA$ reactions and in line with the BUU transport calculation [9].}
(at $p_{K^+} \ge 0.8$ GeV/c
\footnote{It should be noted that for kinematical conditions of our experiment the data points in Figs. 2 and 3 which correspond
to the laboratory kaon momenta $p_{K^+} \ge 0.8$ GeV/c are the subthreshold data points.}
). It is also seen that the inclusion of the $YN$-FSI effects (dash-dotted lines in Figs. 2 and 3) enhances the kaon spectra from the
one-step processes (1), (2) at these momenta by a factor of about 2. As a result, our overall calculations (the sum of contributions
both from the one-step (1), (2) and from the two-step (33), (34) reaction channels, solid lines) reproduce well the measured
$K^+$ invariant differential cross section in the case of $^{9}$Be target nucleus. As for the $^{63}$Cu target nucleus, while our
full model calculations with including the $YN$-FSI effects reproduce quite good the shape of the measured kaon spectrum, they
nevertheless overestimate the strength of it by a factor of around 2. The calculations without including of the FSI effects provide better description
of the data obtained on $^{63}$Cu target.

            In Figures 4 and 5 we show a comparison of the results of our calculations for the double
differential $K^+$ spectra at an angle of 4$^{\circ}$ for $p+^{12}$C collisions at beam energies of 1.5 and 2.0 GeV with the
experimental data from [8] taken by the ANKE-at-COSY Collaboration at forward laboratory angles $\le$ 12$^{\circ}$. These figures demonstrate that the proton--induced
reaction channels (1), (2) dominate the $K^+$ production also in the case of $^{12}$C target nucleus for both bombarding
energies under consideration. It can be also seen from them that while the data at 2.0 GeV incident energy are rather well
described by the overall calculations when taking the $YN$-FSI effects into account (solid line in Fig. 5), the spectrum at
1.5 GeV beam energy is underestimated largely in the momentum range above 300 MeV/c for these
effects included. This is possible tied to the fact that two different normalization methods have been used for the data sets at 2.0 and 1.5 GeV beam energies.

Reasonable description of the kaon spectra in the wide momentum range at different production angles (Figs. 2--5) offers the possibility to compare the data [8] with those obtained in this experiment in spite of the kinematical conditions of the ANKE measurements are not identical to those of the present experiment. The double differential cross sections measured by us at initial proton energy $\epsilon_0$=1.7 GeV
\footnote{The absolute normalization in [8] was not established at the initial proton energy $\epsilon_0$=1.75 GeV which is close to $\epsilon_0$=1.7 GeV in the present study.}
 lie within the confines of the solid curves in Figs. 4, 5 refering to $\epsilon_0$=1.5 GeV and $\epsilon_0$=2 GeV. Moreover, the target mass dependences - presented in the form $\sigma \sim A^{\alpha}$ - are quite similar. The value of cross section ratio Au/C measured at 1.75 GeV beam energy (Table 8 in [8])\footnote{The cross section ratios do not depend on the normalization.} corresponds to the magnitude of $\alpha$=0.47$\pm$0.04 for maximal kaon momentum of 0.596 GeV/c, while the Cu/Be ratio  (see Table 1) yields $\alpha$=0.48$\pm$0.09 for minimal kaon momentum of 0.675 GeV/c. Thus, one can conclude that the results of two measurements are compatible.

       The data from the present experiment for the invariant inclusive cross sections for the production of $K^-$ mesons
at an angle of 10.5$^{\circ}$ in the interaction of 2.25 and 2.4 GeV protons with the $^{9}$Be nuclei are displayed, respectively,
in Figs. 6 and 7 in comparison to the calculated ones by (6), (40), (48). The same as in Fig. 7, but for the interaction of 2.4 GeV
protons with the $^{63}$Cu nuclei is given in Fig. 8. It is seen that the contributions from the resonant channels
$pN \to pN{\phi}$, $pn \to d{\phi}$, $\phi \to K^+K^-$ are small at antikaon momenta above 1.0 GeV/c
\footnote{In particular, this is in line with the conclusion about the relative role of the phi production/decay process
$pN \to pN{\phi}$, $\phi \to K^+K^-$ in "hard" antikaon production also from $p$Be and $p$Cu reactions inferred in [17].}
, whereas at lower $K^-$ momenta their role is essentially more important. Thus, at these momenta the contributions to the $K^-$
production from the production/decay sequence $pN \to pN{\phi}$, $\phi \to K^+K^-$ and pion--induced
reaction channel ${\pi}N \to NKK^-$ and from primary non-resonant proton--nucleon production process
$pN \to NNKK^-$ and resonant channel $pn \to d{\phi}$, $\phi \to K^+K^-$ are comparable. It is also clearly seen that the main
contributions to the subthreshold antikaon production in $p+^{9}$Be, $p+^{63}$Cu interactions at low antikaon momenta
($p_{K^-} \le 0.6$ GeV/c) come from the secondary $K^-$ production processes ${\pi}N \to NKK^-$ and
$pN \to pN{\phi}$, $\phi \to K^+K^-$, whereas at high antikaon momenta ($p_{K^-} \ge 1.0$ GeV/c) the primary non-resonant
(non-$\phi$) proton--induced reaction channels $pn \to dK^+K^-$ and $pN \to NNKK^-$ are dominant.
It should be pointed out that their dominance here for the $^{63}$Cu target nucleus, contrary to the case of $^{9}$Be
nucleus, is less pronounced.
The inclusion of the FSI effects among the outgoing nucleons participating in the non-resonant $pN \to NNKK^-$
channel (dash-dotted lines in Figs. 6--8)
enhances the high-momentum parts ($p_{K^-} \ge 1.0$ GeV/c) of the antikaon spectra from this channel by a factor of about
1.2--2.0
and practically does not affect their low-momentum parts ($p_{K^-} \le 0.6$ GeV/c). As a result, our full calculations
( solid lines in Figs. 6--8)
reproduce quite well the measured
$K^-$ invariant inclusive cross sections for both target nuclei.
The important point is that to reproduce the high-momentum tails of the antikaon spectra it was also significant to take into consideration the contribution from the new $K^-$ production channel non-$\phi$ $pn \to dK^+K^-$.

Good agreement of the measured antikaon spectra with our calculations evidences for weak influence of the in-medium kaon and antikaon potentials on the $K^-$ yields from the reactions with $KK^-$ in the final state $pN \to NNKK^-$, ${\pi}N \to NKK^-$,  $pn \to dKK^-$, $pN \to pN\phi$, $pn \to d\phi$, $\phi \to K^+K^-$. This observation is in line with the analysis [22, 23], based on the spectral function approach, where it has been shown that the simultaneous application of the density-dependent weak repulsive $K^+$ and relatively deep attractive $K^-$ nuclear potentials of +22 MeV and -126 MeV at saturation density remains unaffected  antikaon spectra with momentum of more than 0.8 GeV/c compared to the case with zero kaon and antikaon nuclear mean fields. Similar cancellation has been found in [17] where reasonable description of the $K^-$ excitation function for the $pBe$ and $pCu$ reactions measured for $K^-$ momentum of 1.28 GeV/c at beam energy $<$3 GeV was reached in the frame of the folding model with nucleon momentum distribution (13) assuming vacuum $K^+, K^-$ masses as well as kaon and antikaon masses modified by their potentials. These arguments are not valid for the reactions with the $K^-$ alone in the final state. The impact of the antikaon self-energy on the $K^-$ yield from $YN \to NNK^-$ reaction can be essential.

The importance of the strangeness exchange processes $pN \to K^+YN$; $YN \to NNK^-$, $Y\pi \to NK^-$ for antikaon production in $pA$ reactions was pointed out in [18, 21]. According to the BUU calculation the contribution of the strangeness exchange reactions in the cross sections for antikaon production at 40-56$^0$ by 2.5 GeV protons reaches to 50--60 \% for heavy gold target. No significant influence is found for light carbon target. It should be pointed out that the results of such calculations are model-dependent, since the cross section of the $YN \to NNK^-$ reaction experimentally is not known. To produce $K^-$ at the second step of the strangeness exchange process $YN \to NNK^-$ the hyperon with high momentum exceeding 1.3 GeV/c at the first step $pN \to K^+YN$ is required. All known experimental data on forward hadron production from nuclei with A$\ge$9 evidence for strong decreasing of the cross sections with Feynman variable $X^{R}_{F}$ (see, for instance, Fig. 1). The drop of the high-momemtum parts of the ${\pi}^+$ and ${\pi}^-$ spectra in the energy range from 2 to 24 GeV [17, 49, 50] as well as decrease of the antiproton spectra [14] at initial proton energies of 3-5 GeV follow the dependence $(1-X^{F}_{R})^\delta$ with  ${\delta}\ge$ 4. Thus, it is natural to suppose that the high-momentum part of the hyperon spectrum also follows this dependence  with $\delta{\ge}$4. In Fig. 9 we compare the behavior of the $X^{F}_{R}$ as a function of $P_{t, min}$ for the present experiment and [18]. The $P_{t, min}$ is a minimal (threshold) momentum of intranuclear nucleon required for the $K^-$ production in direct $pN \to K^-KNN$ process in the kinematics of particular experiment. The value of $P_{t, min}$ is defined by eq. 24 with $S_{NNK} = (2m_{N}+m_{K})^2$. Experimental data obtained in [17, 18] as well as data from the present study evidence that at equal $P_{t, min}$ the double  differential cross sections for the antikaon production are comparable. Fig. 9 demonstrates that the values of $X^{F}_{R}$ calculated for the kinematical conditions of our experiment significantly exceed ones for the experiment [18]. Hence, the contribution of the $YN \to NNK^-$ channel in the kinematics of the present experiment is less then that in [18] at least by a factor of 7-10
\footnote{We use $\delta$=3.5 for this estimate}.
Our estimate shows that the contribution of strangeness exchange channel in the cross section for antikaon production on $Cu$ target does not exceed 10-15\%. Good description of the $K^-$ data obtained on both Be and Cu nuclei (see Figs. 6, 7, 8) supports this conclusion.

    Taking into account the above considerations, we conclude that our model calculations support the proposed
mechanisms for the subthreshold and near-threshold charged kaons production in the case of $p+^{9}$Be,  $p+^{12}$C
and  $p+^{63}$Cu collisions.

\section*{4. Conclusions}

\hspace{1.5cm} In this paper we have presented the experimental data on the invariant inclusive cross sections for the
production of $K^+$ and $K^-$ mesons in the momentum range from 0.6 to 1.3 GeV/c  at a lab angle of 10.5$^{\circ}$ in $p+$Be, $p+$Cu interactions, respectively, at
1.7 and 2.25, 2.4 GeV beam energies. This is the first measurement of antikaon spectra in proton--induced reactions on nuclei
in the subthreshold energy regime. The above experimental data as well as those from [8] for $K^+$ production in $p+$C
collisions at 1.5 and 2.0 GeV incident energies, obtained by the ANKE-at-COSY Collaboration, are compared to the results
of calculations in the framework of an appropriate folding model for incoherent primary proton--nucleon, secondary pion--nucleon
kaon and antikaon production processes and processes associated with the creation of antikaons via the decay of intermediate
$\phi$-mesons. The model is based on the struck target nucleon momentum distribution and on free elementary cross sections. The comparison of the data to our model calculations indicates that the
proposed mechanisms for the subthreshold and near-threshold charged kaons production in $pA$ reactions are realistic enough. It was shown for the first time that the antikaon production for the momentum $p_{K}< $0.8 GeV/c is dominated by the $pN \to pN\phi$; $\phi \to K^+K^-$ channel  on light $^{9}Be$ nucleus. On $^{63}Cu$ nucleus the main contribution to the cross section comes from this channel and ${\pi}N \to NKK^-$ process.

It was also shown that the strangeness exchange mechanism plays a minor role in the processes of the production of the antikaons with momentum larger than 0.6 GeV/c emitted in forward direction from proton-induced reactions. We have argued that the cross sections of the reactions with $KK^-$ in the final state are weakly influenced by the in-medium kaon and antikaon mean fields. The performed calculations showed that accounting for the hyperon-nucleon and nucleon-nucleon FSI effects mostly results in better description of the high momentum tails of the kaon and antikaon spectra.

\section*{Acknowledgments}

\hspace{1.5cm} Two of the authors (Yu. K. and E. P.) acknowledge M. Hartmann, which has attracted our attention
to the new possible $K^-$ production channel non-$\phi$ $pn \to dK^+K^-$.
This study was partly supported by the Russian Fund for Basic Research under grant No. 07-02-91565.

\newpage
\begin{figure}[h!]
\centerline{\epsfig{file=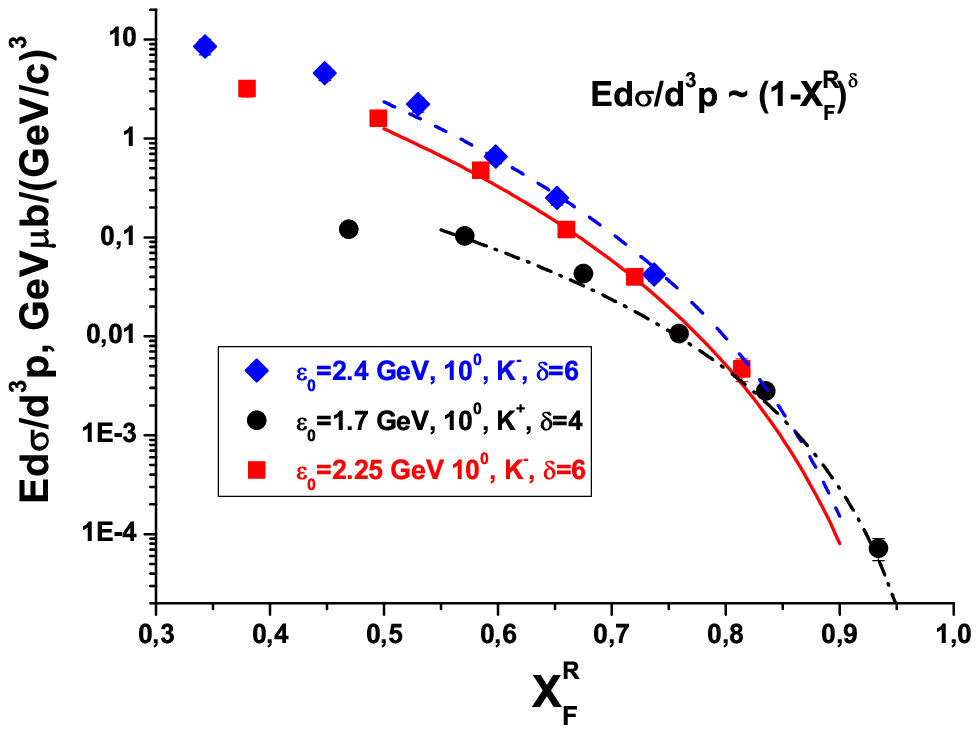,width=.66\textwidth,angle=0,silent=,
clip=}}
\caption{\label{centered}
Invariant cross sections for the kaon and antikaon production as a function of radial Feynman variable $X^{R}_{F}$.}
\end{figure}
\begin{figure}[h!]
\centerline{\epsfig{file=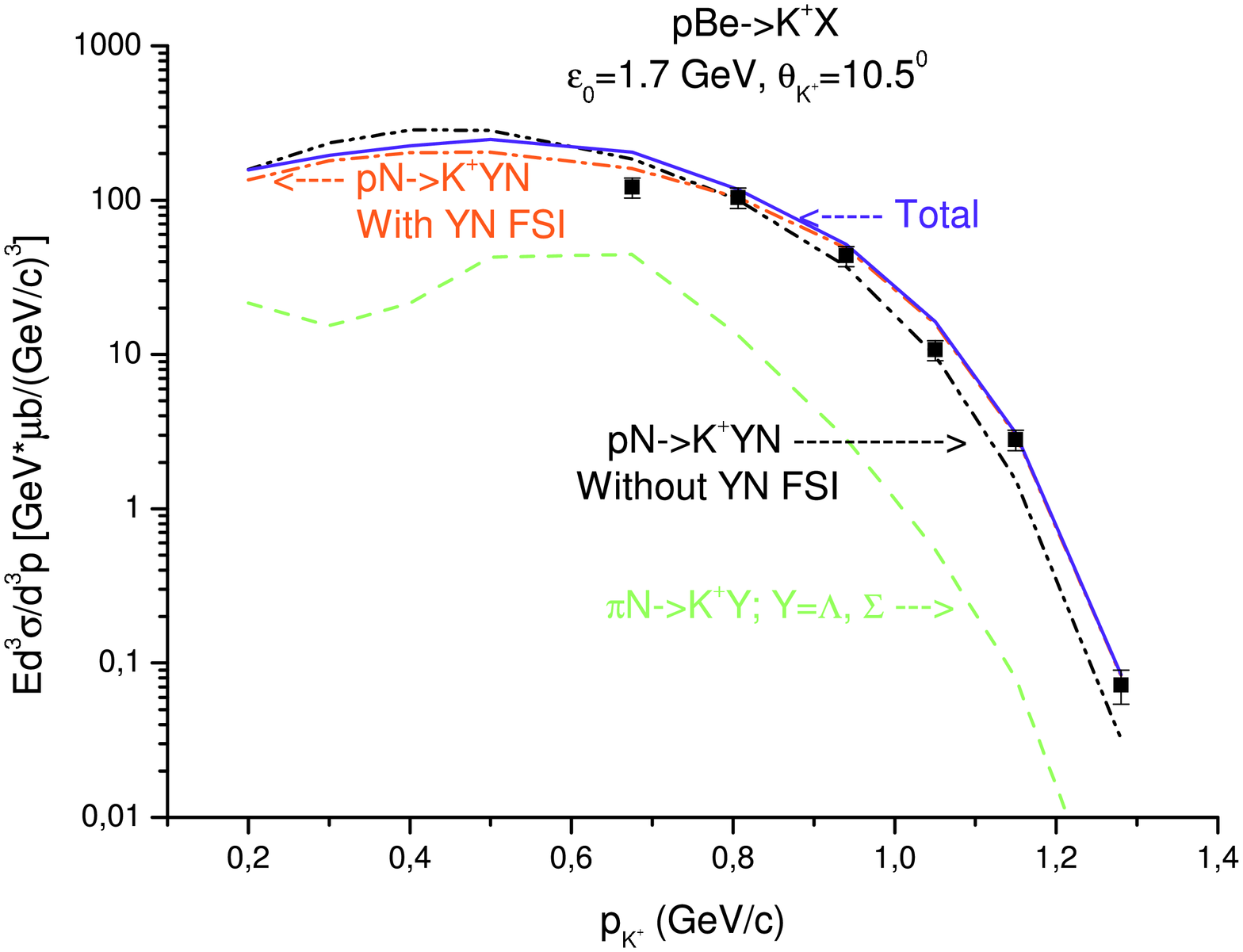,width=.66\textwidth,angle=270,silent=,
clip=}}
\caption{\label{centered}
Invariant cross sections for the production of $K^+$ mesons at an angle of $10.5^{\circ}$ in the
interaction of 1.7 GeV protons with the $^{9}$Be nuclei as functions of kaon momentum. The experimental data (full squares)
are obtained at the ITEP accelerator (see Table 1).
The curves are our calculation.
The dashed lines with one and two dots are calculations for the primary
production processes $pN \to K^+YN$; $Y=\Lambda, \Sigma$ with and without including the FSI effects among the outgoing
hyperons and nucleons. The dashed line is calculation for the secondary production processes ${\pi}N \to K^+Y$
with an intermediate pion. The solid line is the sum of the dashed and dash-dotted lines.}
\end{figure}
\begin{figure}[h!]
\centerline{\epsfig{file=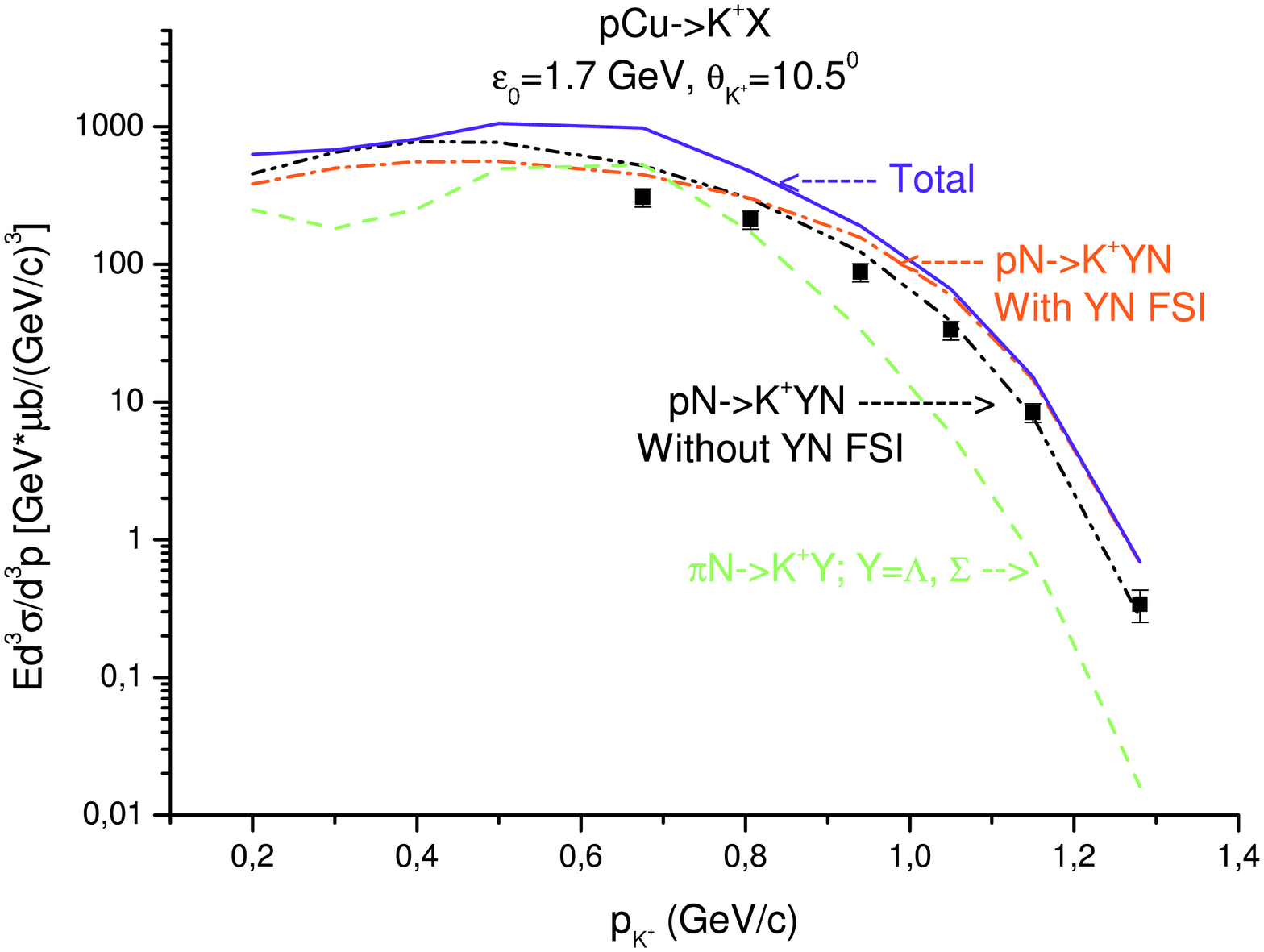,width=.66\textwidth,angle=270,silent=,
clip=}}
\caption{\label{centered}
The same as in Fig. 2, but for the interaction of protons with the $^{63}$Cu nuclei.}
\end{figure}
\begin{figure}[h!]
\centerline{\epsfig{file=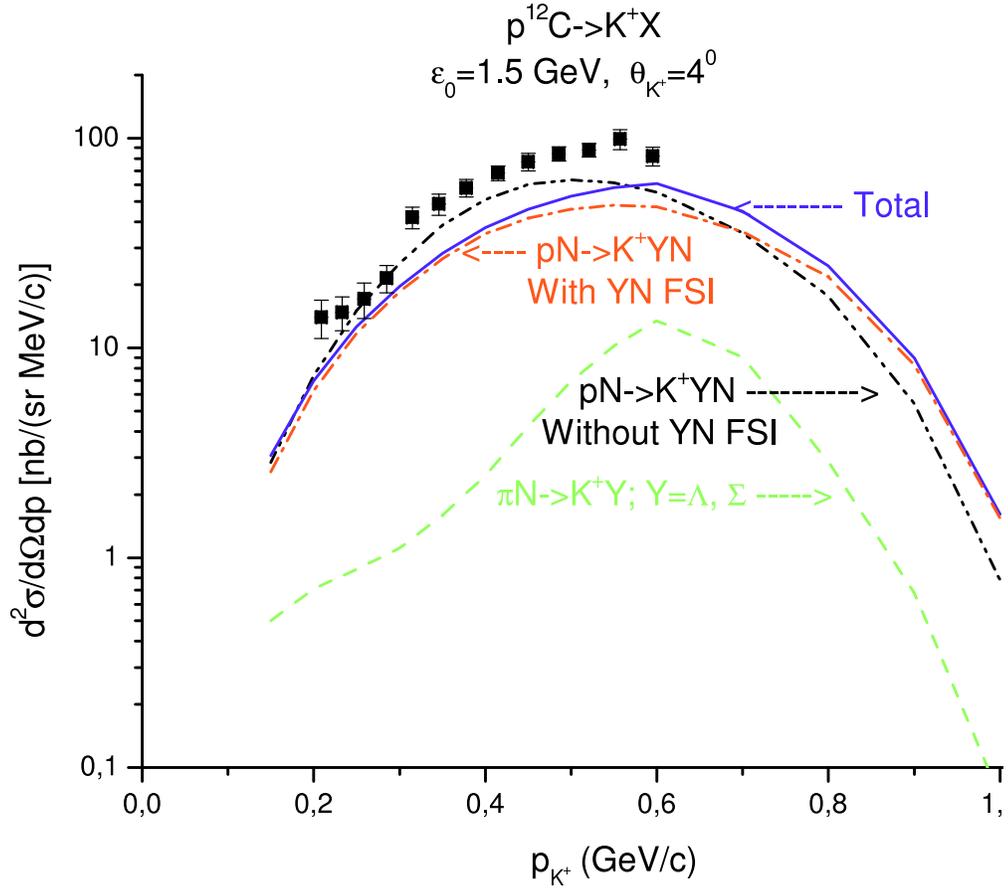,width=.66\textwidth,angle=270,silent=,
clip=}}
\caption{\label{centered}
Double differential cross sections for the production of $K^+$ mesons in the
interaction of 1.5 GeV protons with the $^{12}$C nuclei as functions of kaon momentum. The experimental data (full squares), taken at forward laboratoty angles
$\le$ 12$^{\circ}$,
are from [8]. The curves are our calculation at an angle of $4^{\circ}$. The notation of the curves is identical to that in Fig. 2.}
\end{figure}
\begin{figure}[h!]
\centerline{\epsfig{file=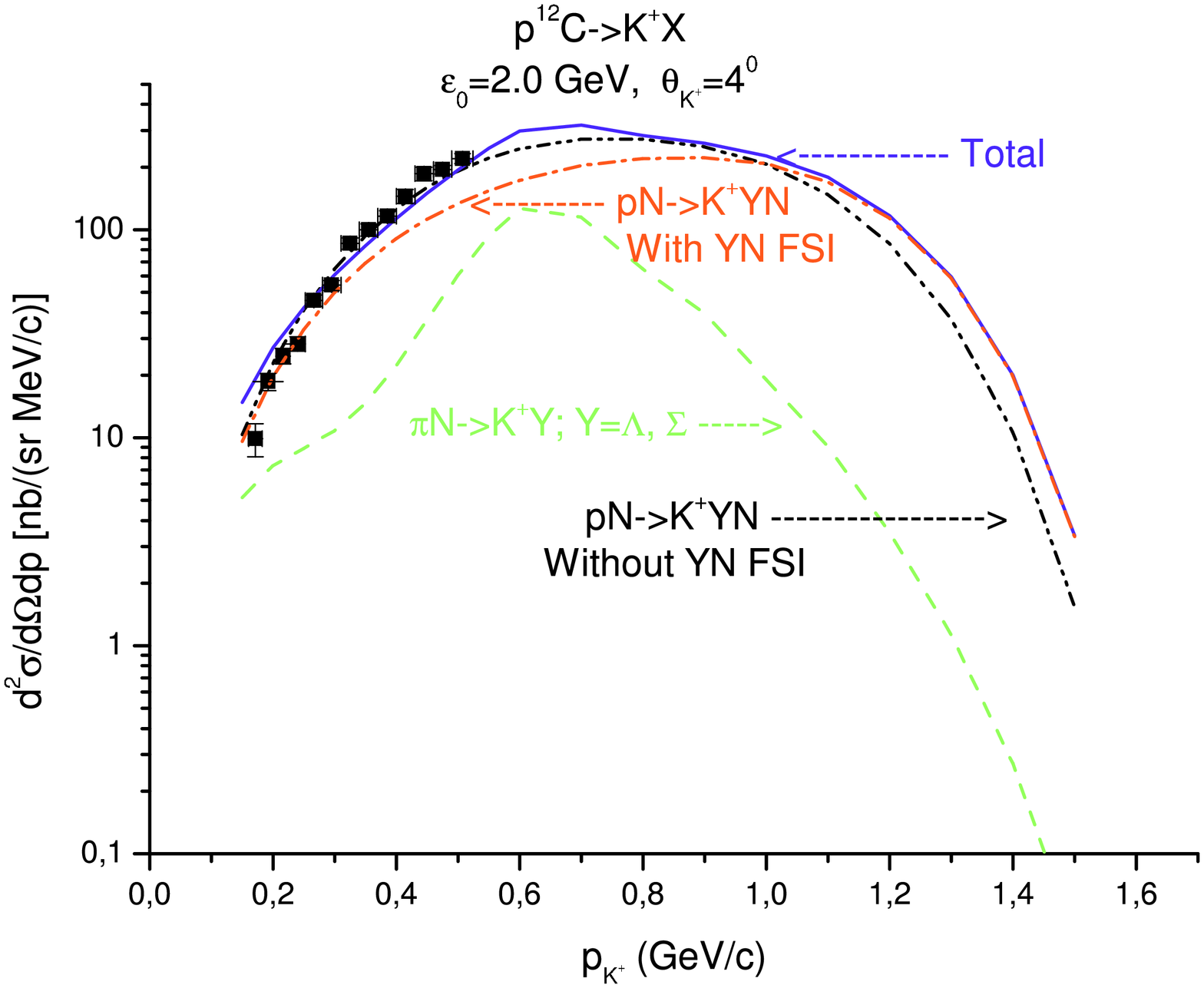,width=.66\textwidth,angle=270,silent=,
clip=}}
\caption{\label{centered}
The same as in Fig. 4, but for 2.0 GeV beam energy.}
\end{figure}
\begin{figure}[h!]
\centerline{\epsfig{file=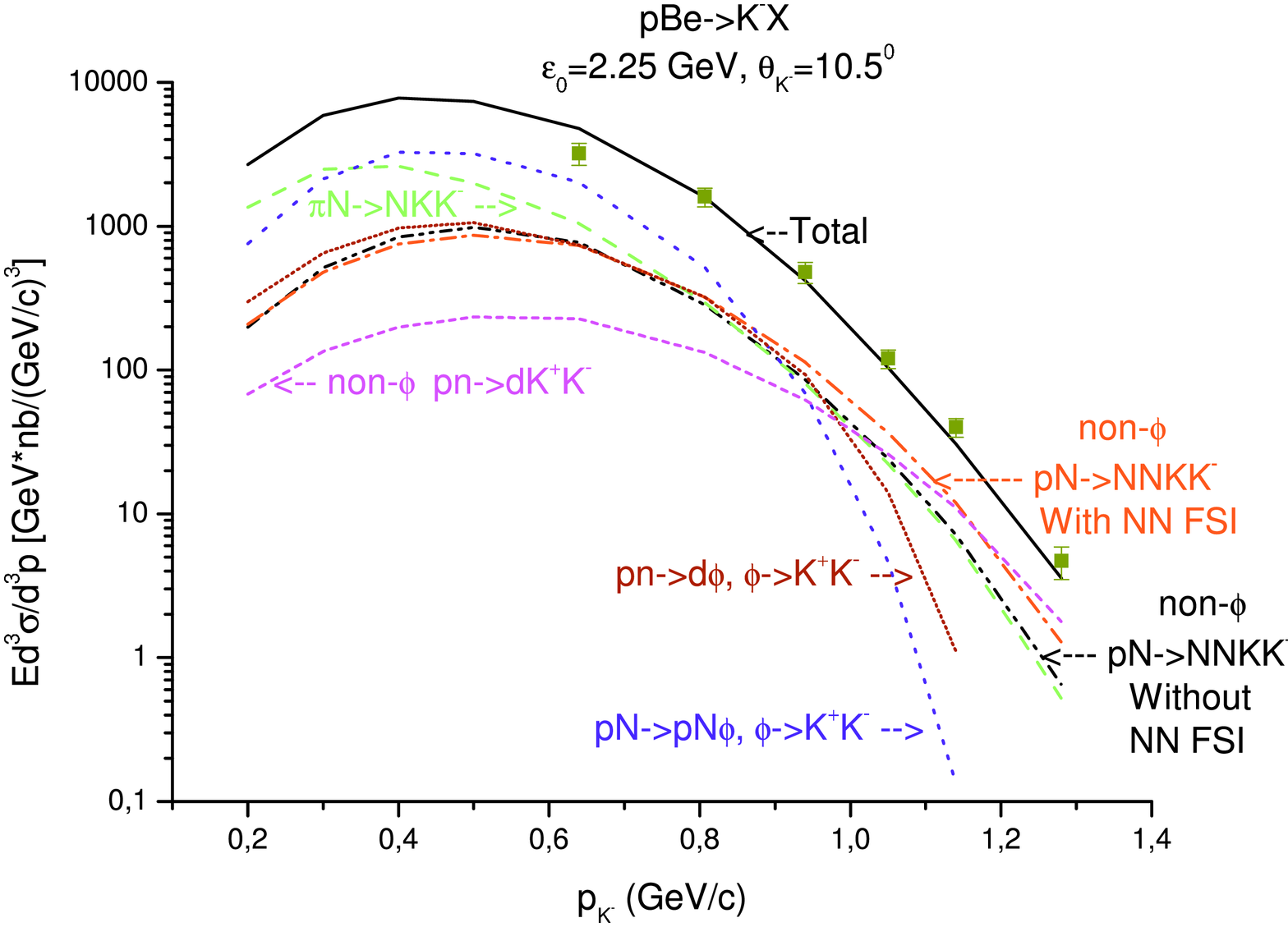,width=.66\textwidth,angle=270,silent=,
clip=}}
\caption{\label{centered}
Invariant cross sections for the production of $K^-$ mesons at an angle of $10.5^{\circ}$ in the
interaction of 2.25 GeV protons with the $^{9}$Be nuclei as functions of antikaon momentum. The experimental data (full squares)
are obtained at the ITEP accelerator (see Table 2).
The curves show our calculation. The dashed lines with one and two dots show the calculation for the primary
production process $pN \to NNKK^-$ with and without including the FSI effects among the outgoing nucleons. The dashed and
short-dashed lines represent the calculation, respectively, for the secondary production process ${\pi}N \to NKK^-$ and direct
non-${\phi}$ $pn \to dK^+K^-$ channel. The dotted and short-dotted lines show the calculation for the secondary creation processes
$pN \to pN{\phi}$, $\phi \to K^+K^-$ and $pn \to d{\phi}$, $\phi \to K^+K^-$. The solid line is the sum of the dash-dotted, dashed,
short-dashed, dotted and short-dotted lines.}
\end{figure}
\begin{figure}[h!]
\centerline{\epsfig{file=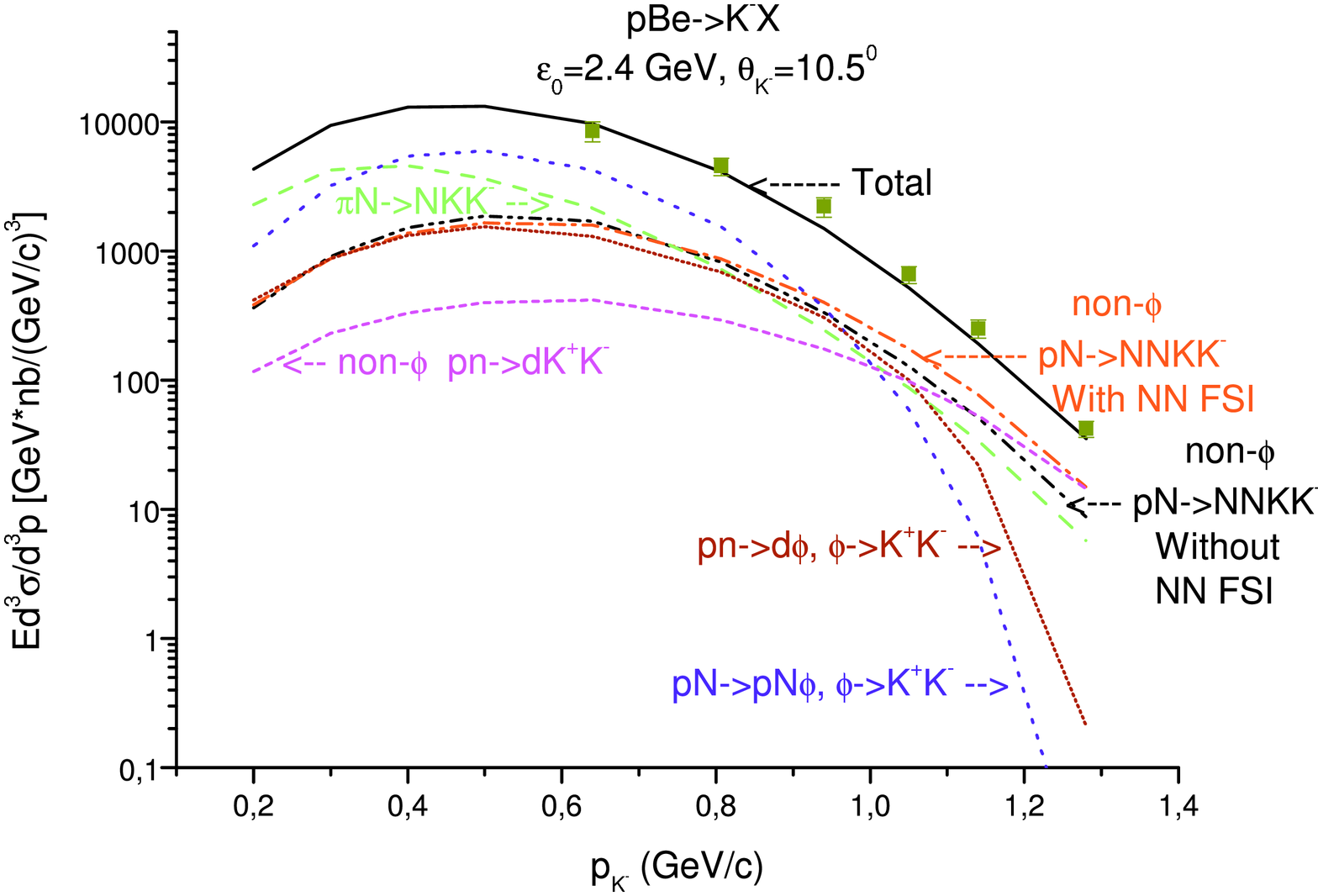,width=.66\textwidth,angle=270,silent=,
clip=}}
\caption{\label{centered}
The same as in Fig. 6, but for 2.4 GeV beam energy.}
\end{figure}
\begin{figure}[h!]
\centerline{\epsfig{file=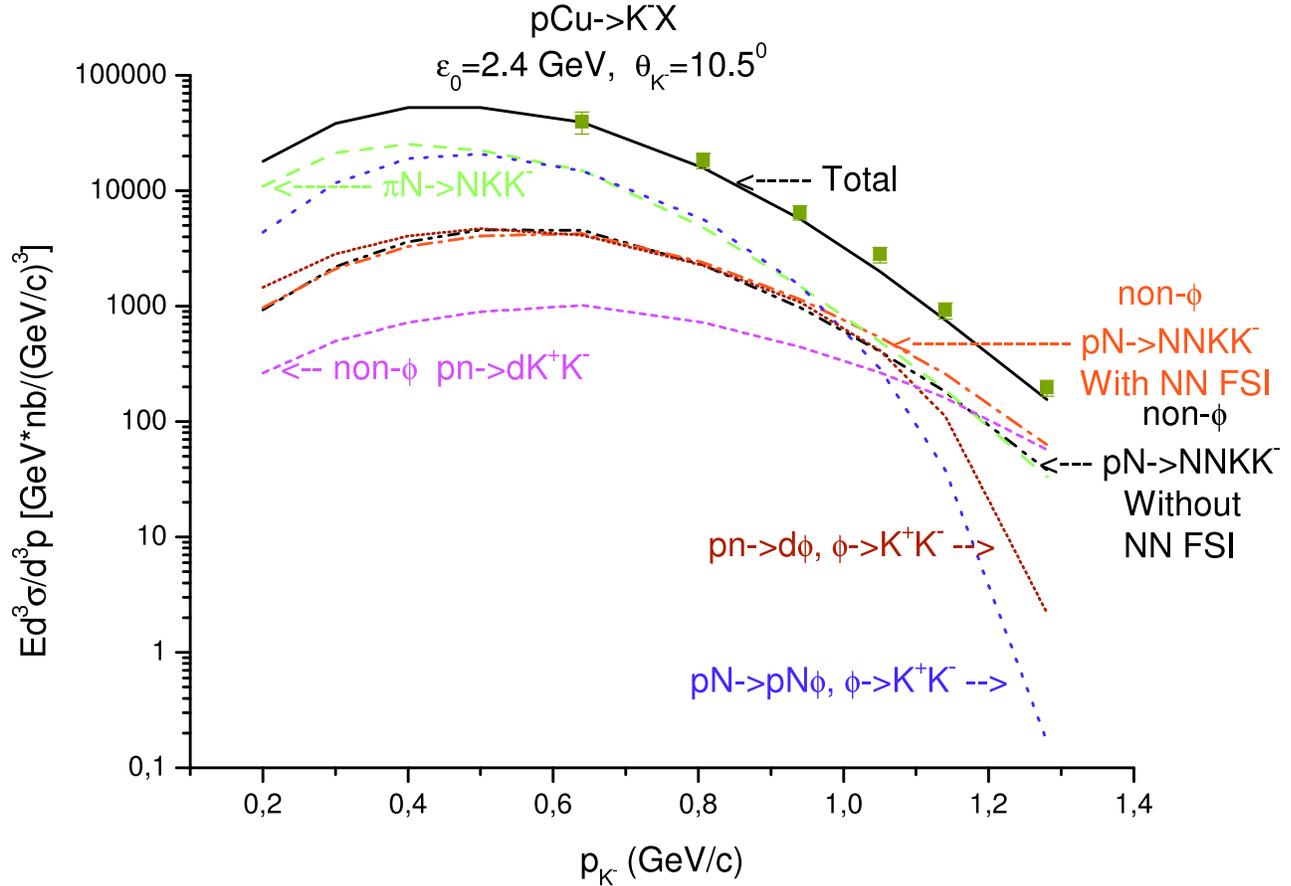,width=.66\textwidth,angle=270,silent=,
clip=}}
\caption{\label{centered}
The same as in Fig. 6, but for the interaction of 2.4 GeV protons with the $^{63}$Cu nuclei.}
\end{figure}
\begin{figure}[h!]
\centerline{\epsfig{file=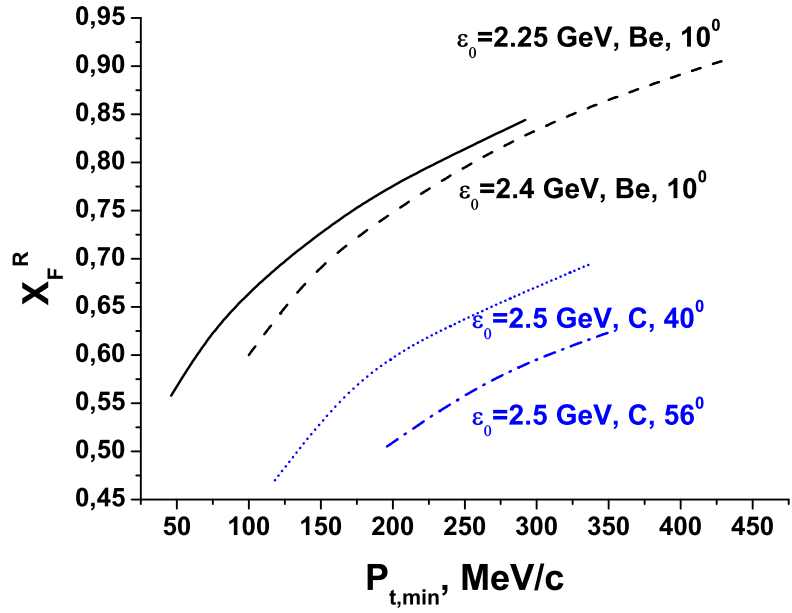,width=.66\textwidth,angle=0,silent=,
clip=}}
\caption{\label{centered}
Radial Feynman variable $X^{R}_{F}$ for the hyperon production as a function of $P_{t, min}$ calculated for the kinematical conditions of the present experiment (solid and dashed curves) and [18] (dotted and dash-dotted curves).}
\end{figure}
\end{document}